\documentclass[12pt]{article}
\begin{document}
\newcommand{\beq}{\begin{equation}}
\newcommand{\eeq}{\end{equation}}
\title{Equivalence of the long-wavelength approximation and the truncated 
Taylor expansion in relativistic Coulomb excitation}
\author{B.F. Bayman\\
{\it School of Physics and Astronomy, University of Minnesota,}\\
{\it 116 Church Street S.E., Minneapolis, MN 55410, U.S.A.}\\
and\\
F.Zardi\\
{\it Istituto Nazionale di Fisica Nucleare and Dipartimento di Fisica,} \\
{\it Via Marzolo,8 I-35131 Padova,Italy.}}
\maketitle
\begin{abstract}
{\bf The long-wavelength approximation and the truncated 
Taylor expansion are frequently used in the theory of relativistic Coulomb 
excitation to obtain multipole expansions of the interaction.
It is shown in this note that these two approximations are exactly 
equivalent.}
\end{abstract}

PACS numbers: 25.70.De, 25.75.-q, 24.30.Cz

\section{Introduction}

Relativistic Coulomb excitation is an important tool for the study of 
several aspects of nuclear structure that cannot be explored through
the nuclear interaction [1-7].
%\cite{WA79,BerBau,EML,ChFr,Bra,Cat,ABE}.
An important quantity in the analysis of relativistic Coulomb excitation
is the retarded scalar potential\cite{Jack}, at point ${\bf r}$ and 
time $t$, due to a
projectile moving with speed $v$ along a straight-line trajectory:
\begin{equation}
\Phi({\bf r},t)=\frac{Z_Pe \gamma}{\sqrt{(x-b_x)^2+(y-b_y)^2+\gamma^2(z-vt)^2}}
\end{equation}
The charge of the projectile is $Z_{P}e$, the impact parameter of the
trajectory is $b$, and the angle between the reaction plane
and the ${\hat x}-{\hat z}$ plane is $\phi_b$ ($b_x=b\cos\phi_b,
~b_y=b\sin\phi_b$). To calculate matrix elements of the
electromagnetic interaction between nuclear angular momentum
eigenstates to be used in coupled-channel time-dependent differential
equations, it would be convenient to have a multipole expansion of
eq.(1), but unfortunately no exact multipole expansion is available for
$\Phi({\bf r},t)$.

Alder and Winther\cite{WA79} have 
derived an exact multipole expansion for the Fourier
transform, $V(\omega,b)$, of $\Phi({\bf r},t)$. This representation can be 
used directly 
to calculate the Born approximation, and has been used in ref.\cite{BZ} 
in a coupled integral equation formulation of the Schroedinger equation. 
However this approach is difficult to implement when many excited 
states have to be included in the calculation. 

Two approximate multipole expansions in the $t$-representation 
have been developed \cite{Bra,Cat}. They start
from two different {\it exact} expansions of $\Phi({\bf r},t)$, and then
truncate these expansions in different ways. The purpose of this report
is to show that the two approximate expansions so obtained are
identical.

\section{ The Long-Wavelength Approximation (LWLA)}.

If we invert the $V(\omega,b)$ transform of Alder and Winther,
we can write a formally exact multipole expansion
\beq
\Phi({\bf r},t)~=~\frac{\hbar}{2
\pi}\int_\infty ^\infty e^{-i\omega t}~
\Bigl[~\sum_{\lambda
\mu}e^{-i\mu \phi_b}C_{\lambda,\mu}(\omega)K_\mu(\frac{|\omega|b}{v \gamma})
j_{\lambda}(\frac{|\omega|}{c}r)Y_{\mu}^{\lambda}({\hat r}~\Bigr])d\omega~,
\eeq
with
$$
C_{\lambda,\mu}(\omega)~\equiv~\frac{2Z_Pe}{\hbar v}~
{\cal G}_{\lambda,\mu}
$$
where
\begin{eqnarray*}
{\cal G}_{\lambda,\mu}&\equiv& \frac{i^{\lambda+\mu}}{(2 
\gamma)^{\mu}}~(\frac{\omega}{|\omega|})^{\lambda-
\mu}~(\frac{c}{v})^{\lambda}
~\sqrt{4 \pi~ (2\lambda+1)~(\lambda-
\mu)!~(\lambda+\mu)!}\\
&\times&\sum_{n}\frac{1}{(2 \gamma)^{2n}(n+\mu)!n!(\lambda-\mu-2n)!}
\end{eqnarray*}
and
$$
j_{\lambda}(\frac{|\omega|}{c}r)=\sum_{\ell=\lambda,\lambda+2,\ldots}
(\frac{|\omega|}{c}r)^\ell~\frac{(-\frac{1}{2})^{\frac{\ell-
\lambda}{2}}}{(\frac{\ell-\lambda}{2})!(\ell+\lambda+1)!!}
$$
If this is substituted into the Alder-Winther expansion, and the coefficient
of the $r^\ell~Y^\lambda_\mu({\hat r})$ term is identified, we can
write
\beq
\Phi({\bf r},t)=\sum_{\lambda,\mu,\ell(=\lambda,\lambda+2,\ldots)}A^{\lambda,
\ell}_{\mu}~r^\ell Y^\lambda_\mu ({\hat r})
\eeq
with
\beq
A^{\lambda,\ell}_{\mu}~=~\frac{\hbar}{2 \pi}\frac{(-\frac{1}{2})^{\frac{\ell-
\lambda}{2}}~e^{-i \mu \phi_b}}{(\frac{\ell-\lambda}{2})!(\ell+\lambda+1)!!}
~\times~ \int_{-
\infty}^{\infty}e^{-i\omega t}(\frac{|\omega|}{c})^\ell C_{\lambda,\mu}(\omega)
K_\mu(\frac{|\omega|b}{v \gamma})~d\omega
\eeq
The LWLA consists of keeping only the first term in the expansion of the
spherical Bessel function, which is equivalent to the approximation
\begin{equation}
\Phi({\bf r},t)\sim \sum_{\lambda,\mu}A^{\lambda,
\lambda}_{\mu}~r^\lambda Y^\lambda_\mu ({\hat r})~~~~~~~~~{\rm LWLA}.
\end{equation}
This is the approximation used, e.g., in ref.\cite{Cat}.

\section{The Truncated Taylor Expansion (TTEA)}

Another formally exact expansion of $\Phi({\bf r},t)$ is a Taylor expansion,
which can be written 
$$
\Phi({\bf r},t)=
\sum_{\ell=0}^{\infty}\frac{1}{\ell!}({\bf r \cdot \nabla_s})^\ell
\Phi({\bf s},t)|_{{\bf s}=0}
$$
To express this as a multipole expansion, we write the dot product in
terms of Legendre polynomials and spherical harmonics. For any vectors
${\bf a,b}$
$$
({\bf a \cdot b})^\ell=a^\ell b^\ell (\cos(\omega_{ab}))^\ell
=a^\ell b^\ell \sum_{\lambda=\ell,\ell-2,\ldots}\frac{(2
\lambda+1)2^\lambda
\ell! (\frac{\ell+\lambda}{2})!}{(\ell+\lambda+1)!(\frac{\ell-
\lambda}{2})!}~P_{\lambda}(\cos(\omega_{ab}))
$$
$$
=a^\ell b^\ell \sum_{\lambda=\ell,\ell-2,\ldots}\frac{(2
\lambda+1)2^\lambda
\ell! (\frac{\ell+\lambda}{2})!}{(\ell+\lambda+1)!(\frac{\ell-
\lambda}{2})!}(a^2)^{\frac{\ell-\lambda}{2}}(b^2)^{\frac{\ell-
\lambda}{2}}
\times \frac{4 \pi}{2 \lambda+1}\sum_{\mu=-\lambda}^{\lambda}{\cal
Y}^{\lambda}_{\mu}({\bf a}){\cal Y}^{\lambda *}_{\mu}({\bf b}),
$$
with ${\cal Y}^{\lambda}_{\mu}({\bf a})\equiv a^\lambda
Y^{\lambda}_{\mu}({\hat a})$, a homogeneous polynomial of degree 
$\lambda$ in $a_x,a_y,a_z$. If we identify ${\bf a}$ with ${\bf
r}$ and ${\bf b}$ with ${\bf \nabla_{s}}$, we can write
$$
({\bf r \cdot \nabla_s})^\ell=
\sum_{\lambda=\ell,\ell-2,\ldots}\frac{(2
\lambda+1)2^\lambda
\ell! (\frac{\ell+\lambda}{2})!}{(\ell+\lambda+1)!(\frac{\ell-
\lambda}{2})!}(r^2)^{\frac{\ell-\lambda}{2}}
\times \frac{4 \pi}{2 \lambda+1}\sum_{\mu=-\lambda}^{\lambda}{\cal
Y}^{\lambda}_{\mu}({\bf r}){\cal Y}^{\lambda *}_{\mu}({\bf \nabla_s})
(\nabla_s^2)^{\frac{\ell-\lambda}{2}},
$$
so that
$$
\Phi({\bf r},t)
~=~4\pi
\sum_{\lambda=0}^{\infty}\sum_{\ell=\lambda,\lambda+2,
\ldots}\frac{2^\lambda
 (\frac{\ell+\lambda}{2})!}{(\ell+\lambda+1)!(\frac{\ell-
\lambda}{2})!}(r^2)^{\frac{\ell-\lambda}{2}}\nonumber\\
\times\sum_{\mu=-\lambda}^{\lambda}{\cal Y}^{\lambda}_{\mu}({\bf r})
 {\cal Y}^{\lambda *}_{\mu}({\bf \nabla_s})
(\nabla_s^2)^{\frac{\ell-\lambda}{2}}
~\Phi({\bf s},t)|_{{\bf s}=0}~.
$$
Now if we extract the coefficient of $r^\ell~Y^\lambda_\mu({\hat r})$,
we can write
\begin{equation}
\Phi({\bf r},t)=\sum_{\lambda,\mu,\ell(=\lambda,\lambda+2,\ldots)}B^{\lambda,
\ell}_{\mu}~r^\ell Y^\lambda_\mu ({\hat r}),
\end{equation}
with 
\beq
B^{\lambda,\ell}_{\mu}=
4 \pi
\frac{2^\lambda
 (\frac{\ell+\lambda}{2})!}{(\ell+\lambda+1)!(\frac{\ell-
\lambda}{2})!}
 {\cal Y}^{\lambda *}_{\mu}({\bf \nabla_s})
(\nabla_s^2)^{\frac{\ell-\lambda}{2}}
\Phi({\bf s},t)|_{{\bf s}=0},
\eeq
where $\ell=\lambda,\lambda+2,\ldots$. 

Since both Equations (4) and (6) hold identically in ${\bf r}$, it must 
be that 
\beq
A^{\lambda,\ell}_{\mu}=B^{\lambda,\ell}_{\mu},
\eeq
which implies that 
$$
\int_{-\infty}^{\infty}e^{-i\omega t}(\frac{|\omega|}{c})^\ell 
C_{\lambda,\mu}(\omega)
K_\mu(\frac{|\omega|b}{v \gamma})~d\omega
=~e^{i \mu \phi_b}~
(-\frac{1}{2})^{\frac{\ell-
\lambda}{2}}\frac{8 \pi^2}{\hbar}~\left[~{\cal Y}^{\lambda *}_{\mu}({\bf 
\nabla_s})(\nabla_s^2)^{\frac{\ell-\lambda}{2}}~\right]\Phi({\bf s},t)
|_{{\bf s}=0}
$$
$$
=~e^{i \mu \phi_b}~
(-\frac{1}{2})^{\frac{\ell-
\lambda}{2}}\frac{8 \pi^2}{\hbar}~\left[~{\cal Y}^{\lambda *}_{\mu}({\bf 
\nabla_s})(\nabla_s^2)^{\frac{\ell-\lambda}{2}}~\right]
\frac{Z_{{\rm P}}e
\gamma}{\sqrt{(s_x-b_x)^{2}+(s_y-b_y)^{2}
+\gamma^2(s_z-vt)^2}}|_{{\bf s}=0}.
$$

So far we have dealt with the full Taylor expansion. The truncation
introduced by Bertulani {\it et al} \cite{Bra} was to keep only the terms in
Equation (6) in which $\ell=\lambda$. 
This implies the approximation
\begin{equation}
\Phi({\bf r},t)\sim \sum_{\lambda,\mu}B^{\lambda,
\lambda}_{\mu}~r^\lambda Y^\lambda_\mu ({\hat r})~~~~~~~~~{\rm TTEA}.
\end{equation}
Note that in the non-relativistic
limit, in which $\gamma \sim 1$, we have
$$
 \nabla_{\bf s}^2\frac{1}{\sqrt{(s_x-b_x)^{2}+(s_y-b_y)^{2}
+(s_z-vt)^2}}=0
$$
if ${\bf s}=0$ and $b \ne 0$. In this case, all the
$B^{\lambda,\ell}_{\mu}$ vanish except those with $\ell=\lambda$. But if 
$\gamma>1$
$$
 \nabla_{\bf s}^2\frac{1}{\sqrt{(s_x-b_x)^{2}+(s_y-b_y)^{2}
+\gamma^2(s_z-vt)^2}}\ne 0
$$
and we no longer have the $\ell=\lambda$ selection rule. Thus the
$\ell=\lambda$ truncation introduced by Bertulani {\it et al} is a
significant extra assumption.

\section{Equivalence of the LWLA and TTEA}

We have seen that LWLA and TTEA both make exactly the same truncation
($\ell=\lambda$) in the expansion of $\Phi({\bf r},t)$ in terms of
$r^\ell Y^\lambda_\mu ({\hat r})$. Therefore the two approximations are
equivalent.

Since the vector potential associated with $\Phi({\bf r},t)$ is
${\bf A}({\bf r},t)=\frac{v}{c}\Phi({\bf r},t){\hat z}$, the equivalence of 
LWLA and TTEA for the scalar potential implies that these approximations 
are also equivalent when applied to the vector potential.

The merit of the TTEA is that it directly yields explicit expressions 
for $\Phi({\bf r},t)$ and ${\bf A}({\bf r},t)$. However, it is important
to keep in mind that results calculated with these expressions have
exactly the same range of validity as results calculated with the LWLA.
\bigskip
\bigskip
\bigskip

{\bf ACKNOWLEDGEMENTS}
\smallskip

One of us (B.F.B.) acknowledges financial support from INFN

\end{document}